\documentclass[useAMS,usenatbib]{mn2e}
\usepackage[dvips]{graphics}

\title[Two-dimensional hydrodynamic simulation of
an accretion flow with radiative cooling $\ldots$
]{Two-dimensional hydrodynamic simulation of
an accretion flow with radiative cooling 
in a close binary system}
\author[Jun'ichi Sato, Keisuke Sawada and Naofumi Ohnishi]
{Jun'ichi Sato$^{1}$
\thanks{E-mail:junichi@cfd.mech.tohoku.ac.jp (JS)}, 
Keisuke Sawada$^{1}$\thanks{E-mail:sawada@cfd.mech.tohoku.ac.jp (KS)} 
and Naofumi Ohnishi$^{1}$\thanks{E-mail:ohnishi@cfd.mech.tohoku.ac.jp (NO)}
\\
$^{1}$Department of Aeronautics and Space Engineering, Tohoku University, 
Sendai 980-8579, Japan}
\begin{document}

\date{in original form 2002 September 20}

\pagerange{\pageref{firstpage}--\pageref{lastpage}} \pubyear{2002}

\maketitle

\label{firstpage}

\begin{abstract}
Two-dimensional numerical simulations of an accretion flow 
in a close binary system are performed by solving the Euler equations 
with radiative transfer. 
In the present study, the specific heat ratio is assumed to be constant 
while radiative cooling effect is included as a non-adiabatic process. 
The cooling effect of the disc is considered by discharging energy 
in the vertical directions from the top and bottom surfaces of the disc. 
We use the flux-limited diffusion approximation to calculate 
the radiative heat flux values. 
Our calculations show that a disc structure appears 
and the spiral shocks are formed on the disc. 
These features are similar to that observed 
in the case of an adiabatic gas 
with a lower specific heat ratio, $\gamma=1.01$. 
It is found that when radiative cooling effect is accounted for, 
the mass of the disc becomes larger than that assuming $\gamma=5/3$, 
and smaller than that assuming $\gamma=1.01$. 
It is concluded that employing an adiabatic gas 
with a lower specific heat ratio is almost a valid assumption 
for simulating accretion disc with radiative cooling effect.
\end{abstract}


\section{Introduction}\label{section1}
The angular momentum loss of the gas in the accretion disc 
in a close binary system is one of the very important phenomena 
in the astrophysics. The most accepted model 
that explains the mechanism of angular momentum transport is 
the $\alpha$ disc model \citep{SS73,Pringle81}. 
In this model, viscosity originating from turbulence, 
magnetism or whatever, is supposed to transport the angular momentum. 
However, in this model the details of how the viscosity is generated 
are still unknown.

An alternative model that needs no viscosity 
for angular momentum transfer is the spiral shock model 
in which spiral shaped shock waves 
are formed in the accretion disc due to the tidal force 
of the mass losing star. 
This model was proposed by \citet{SMH86a}. 
They conducted two-dimensional hydrodynamic calculations 
of a flow of invisicid adiabatic gas, 
and found the spiral shocks numerically. 
Since then a number of two-dimensional simulations have been carried out, 
and it has been claimed that the spiral shocks really appear 
in the accretion discs 
\citep{SMH86b,Sawada+87,Spruit+87,Larson88,Spruit89,
RS89,Matsuda+90,SPL94}. 
Besides, \citet{Spruit87} obtained self-similar solutions 
of spiral shocks for idealized accretion discs. 

For this spiral shock model, two major points have been questioned 
as regard to the validity of the model:

\begin{description}
\item[(i)] The three-dimensional effect might suppress 
the formation of spiral shocks 
\citep{Lin89,LPS90a,LPS90b,LP93}. 
\item[(ii)] The cooling effect due to radiation is 
not taken into account explicitly. 
Though such effect has been considered 
by assuming a lower specific heat ratio for adiabatic gas, 
the obtained disc was still too hot. 
\end{description}

As regard to the question (i), 
three-dimensional simulations have been carried out mostly 
by particle methods. 
Such simulations tend to give rather diffusive results 
if a sufficient number of particles are not available. 
Spiral shocks are likely disappeared particularly 
for those cases with a lower specific heat ratio 
for which the pitch angle of the spiral shock decreases. 
For example, \citet{YBM97} carried out three-dimensional simulations 
using smoothed particle hydrodynamics method. 
As many as $(5-6) \times 10^{4}$ particles were employed 
in the calculations. 
They showed that an accretion disc was formed and 
spiral shocks appeared for the case of $\gamma=1.2$, 
but spiral shocks disappeared 
for the case of $\gamma=1.1$ and $\gamma=1.01$.

Three-dimensional simulation 
using either the finite difference or finite volume method was 
first carried out by \citet{SM92}. 
They calculated the case of $\gamma=1.2$ 
with a mass ratio of unity using the upwind TVD Roe scheme 
and a generalized curvilinear coordinate system. 
Although the calculation was continued up to only a half revolution period 
because of the limitation of CPU time, 
they successfully showed the existence of spiral shocks 
on the accretion disc. 

\citet{Bisikalo+97a}, \citet{Bisikalo+97b}, \citet{Bisikalo+98a}, 
\citet{Bisikalo+98b} and \citet{Bisikalo+98c} 
carried out three-dimensional numerical simulations of accretion discs 
by means of the finite difference method. 
They employed a TVD Roe scheme with a monotonic flux limiter 
of the Osher's form. 
A Cartesian coordinate system was used in the calculations. 
Their results showed that no hot spot was formed. 
Furthermore, disc formation was inhibited 
for higher $\gamma$ and no spiral shock was seen. 

\citet{MMM00} carried out two- and three-dimensional 
numerical simulations of an accretion disc 
in a close binary system with a higher resolution 
using Simplified Flux vector Splitting finite volume method 
\citep{Jyounouchi+93,SJ94}. 
Their computational region only covered the vicinity 
of the mass accreting star. 
The gas from the mass losing star was assumed to flow 
into the computational domain through a rectangular hole located 
at the L1 Lagrangian point. 
They obtained quasi steady solutions, 
and confirmed the existence of both an accretion disc and spiral shocks 
for all cases with $\gamma=1.2$, $1.1$, $1.05$ and $1.01$. 
It was shown that a smaller $\gamma$ value resulted 
in a more tightly wound spiral shock in two-dimensional calculations, 
but such tendency was not so obvious in three-dimensional cases. 
It was also shown that the spiral shock waves disappeared 
when tidal force due to the mass losing star was artificially cut off. 

\citet{fujiwara+01} carried out three-dimensional simulations 
of a close binary system containing 
both the mass accreting star and the mass losing star, 
and investigated the interaction 
between the L1 stream and the accretion disc. 
In the calculations, the ratio of constant specific heat was chosen 
as $\gamma=1.2$ and $1.01$. 
They confirmed the existence of accretion disc as well as spiral shocks. 
No hot spot was found in the accretion disc. 
Instead, they found that a bow shock wave was formed 
due to the collision of the L1 stream and the rotating disc. 
It was shown that this bow shock wave heated 
the outer part of the accretion disc, 
and also enhanced the density perturbation 
in the disc resulting in a more effective transfer of angular momentum 
by the tidal torque. 

From above numerical results, one can assume the existence 
of spiral shocks rather firmly 
even for three-dimensional accretion discs. 
This view is further supported by a recent observational evidence. 
\citet{SHH97} found the first convincing evidence for spiral structure 
in the accretion disc of the eclipsing dwarf nova binary IP Pegasi 
using the technique known as Doppler tomography. 

Next, let us examine the question (ii). 
In the past numerical simulations of accretion discs, 
it has been customary to assume an adiabatic gas 
with a lower specific heat ratio than that of a mono-atomic gas ($=5/3$) 
to account for non-adiabatic process of radiative cooling. 
So far, no simulation that accounts for radiative cooling effect 
with $\gamma=5/3$ has been made to show 
whether a disc structure is really formed and spiral shocks appear. 
It is also yet to know how good is the approximation 
to employ an adiabatic gas with a lower $\gamma$ 
for simulating actual flowfield with radiative cooling effect. 

Recently, \citet{Stehle99} carried out 
two-dimensional hydrodynamic calculations of accretion discs 
including both the $\alpha$-type viscosity and 
the effect of energy loss from the surface by radiation. 
They followed the evolution of the local disc thickness 
in the one-zone model of \citet{SS99}. 
The computational domain, however, 
only covered the vicinity of the mass accreting star. 
In their calculation, spiral shocks were clearly observed 
which dominated the disc evolution in a hydrodynamical time-scale. 

The purpose of this work is first to obtain the two-dimensional flowfield 
in a close binary system with radiative cooling effect, 
and then to explore whether a disc structure is really formed 
and spiral shocks appear in the accretion disc. 
Moreover, we would like to examine how good is the approximation 
of using a lower $\gamma$ value 
for simulating non-adiabatic processes of radiative cooling 
in the hydrodynamic calculation of accretion discs. 
We therefore solve the two-dimensional Euler equations 
with a constant specific heat ratio of mono-atomic gas 
($\gamma=5/3$) 
for a thin disc on the equatorial plane, 
and include radiative cooling effect by discharging energy 
in the vertical directions from the top and bottom surfaces of the disc. 
We employ the flux-limited diffusion (FLD) approximation 
to obtain radiative heat flux values. 
In the present study, we consider a realistic case 
of the observed close binary system. 

This paper is organized as follows. 
In \S \ref{section2} we briefly describe the numerical method, 
physical assumptions and numerical treatment of radiative transfer. 
In \S \ref{section3} we show the obtained results for accretion disc 
with radiative cooling. 
A detailed comparisons are made with the results 
of assuming an adiabatic gas with lower $\gamma$ values. 
In \S \ref{section4} concluding remarks will be given. 

\section[]{ASSUMPTIONS AND METHOD OF CALCULATIONS}\label{section2}

We consider a SU UMa type of CV: OY Car 
as a realistic close binary system \citep{RK95}. 
The separation between the centers of these two stars 
is $4.22 \times 10^{10} [\mbox{cm}]$. 
The orbital period is $1.55 [\mbox{hour}]$. 
The mass of these two stars are 
$M_{1}=0.69M_{\odot}$ for the mass accreting star 
and $M_{2}=0.07M_{\odot}$ for the mass losing star, respectively. 
In the present study, we assume the surface density 
of the mass losing star 
to be $\rho_{0}=0.5 \times 10^{-9} [\mbox{g cm}^{-3}]$. 

\subsection{BASIC EQUATIONS AND PHYSICAL CONDITIONS}
The basic equations describing a two-dimensional flow of perfect gas 
in a rotating frame of reference can be described as 
\citep{SMH84}
\begin{eqnarray}
&&
\frac{\partial \tilde{Q}}{\partial t}+
\frac{\partial \tilde{F}}{\partial x}+
\frac{\partial \tilde{G}}{\partial y}+
\tilde{H}=0, 
\label{eq-1}
\\
&&
\tilde{Q}=
\left(
\begin{array}{c}
\rho \\
\rho u \\
\rho v \\
e
\end{array}
\right)
,\quad
\tilde{F}=
\left(
\begin{array}{c}
\rho u \\
\rho u^{2}+p \\
\rho uv \\
(e+p)u
\end{array}
\right)
,
\nonumber \\
&&
\tilde{G}=
\left(
\begin{array}{c}
\rho v \\
\rho uv \\
\rho v^{2}+p \\
(e+p)v
\end{array}
\right)
,
\nonumber \\
&&
\tilde{H}=
\left(
\begin{array}{c}
0 \\
\rho f_{x} \\
\rho f_{y} \\
\rho(uf_{x}+vf_{y})+\frac{\partial F_{z}}{\partial z}
\end{array}
\right),
\label{eq-2}
\end{eqnarray}
where $\rho$, $u$, $v$, $e$, $p$, $f_{x}$, $f_{y}$ and $F_{z}$ 
are the density, the $x$ and $y$ components of the velocity, 
the total energy per unit volume, the gas pressure, 
the $x$ and $y$ components of the net force 
that includes the gravity force, 
the Coriolis force and the centrifugal force, 
and the $z$ component of the radiative energy flux, respectively. 
We ignore the radiation pressure because it is negligibly small 
if compared with the gas pressure. 

The equation of state is given by
\begin{equation}
p=\left(\gamma-1\right)
\left\{e-\frac{\rho}{2}\left(u^{2}+v^{2}\right)\right\},
\label{eq-3}
\end{equation}
which is that of an ideal gas characterized
by a constant ratio of specific heats $\gamma$.

\begin{figure*}
\vspace{15cm}
\caption{The generalized curvilinear coordinate used here. 
         The mesh size is 
         $20 \times 61$ in the mass losing star region, 
         $3 \times 19$ in the connecting region and 
         $44 \times 61$ in the mass accreting star region.}
\includegraphics{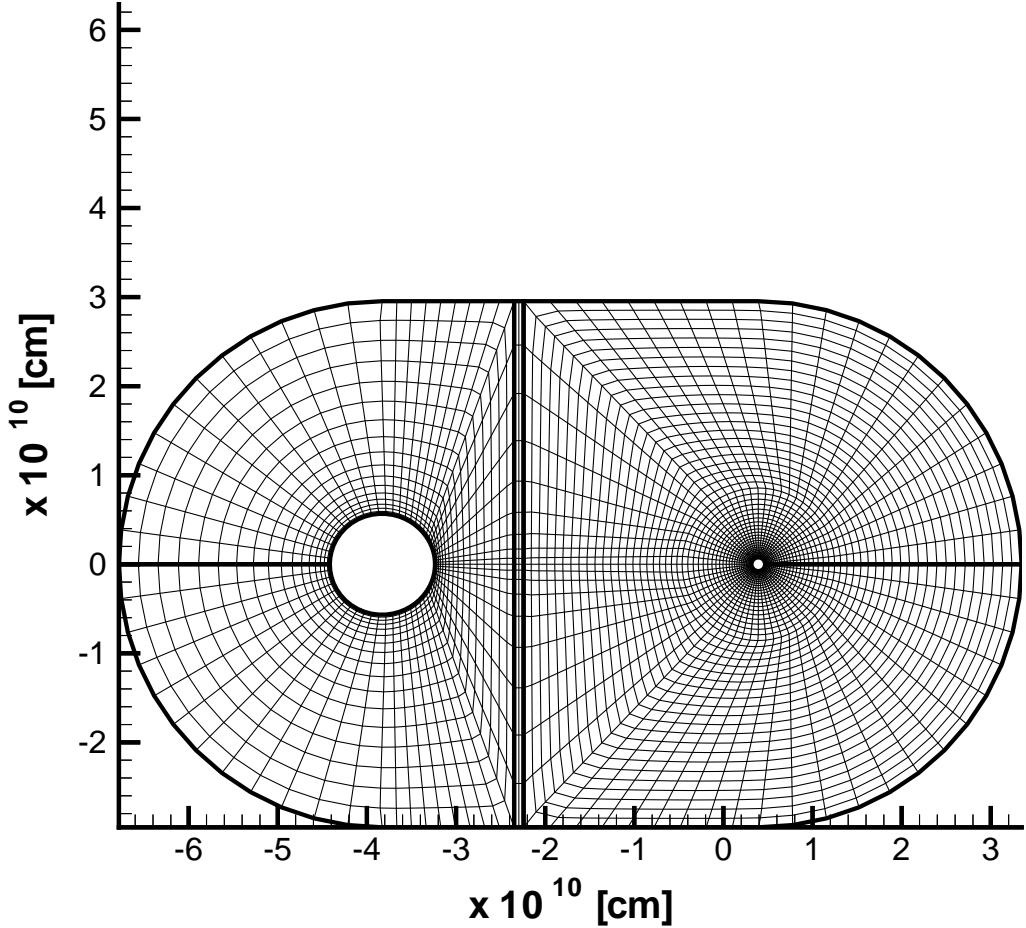}
 \label{fig_1}
\end{figure*}

A finite volume approach is used to discretize the governing equations. 
The AUSM-DV scheme is used to obtain 
the numerical convective flux \citep{WL94}. 
In the time integration, a matrix free LU-SGS method is 
employed \citep{Nakahashi+99}. 
Inner iterations are made to assure the second order temporal accuracy 
in the implicit integration \citep{YK96}. 
A CFL number of $1000$ is assumed in the calculations. 

The parameters that characterize the gas ejected from 
the mass losing component are the sound velocity, 
which is chosen to be $c_{0}=73[\mbox{km s}^{-1}]$, 
and the velocity of the gas inside of the star, 
which is $u_{0}=0$. Although we assume a stationary gas, 
the higher pressure inside of the mass losing star drives the gas flown 
into the computational domain. 
Note that if we employ $A\Omega$ as a typical velocity scale as was 
in the previous works, where $A$ is the separation of the two stars 
and $\Omega$ is the angular velocity of the system, 
the dimensionless sound velocity of the ejected gas becomes $0.15$. 
As a boundary condition at the surface of the mass accreting star 
as well as at the outer boundary of the computational domain, 
we assume a vacuum condition in which any gas reaching 
there is simply absorbed in. 
As an initial condition, we assume a gas with very low density, 
$0.5 \times 10^{-14}[\mbox{g cm}^{-3}]$, 
and high temperature, $3.22 \times 10^{6}/\gamma[\mbox{K}]$, 
that fills up the entire computational domain. 

In order to avoid the numerical instability at the initial stage, 
we first calculate the flow of the gas with $\gamma=5/3$ without 
radiative cooling up to $10^{5}$ time steps using first-order-accurate 
scheme. 
After that we switch to second-order-accurate scheme. 
At the same time, the $\gamma$ value is changed for relevant cases. 
The radiative cooling effect is also included for non-adiabatic cases. 

Our computational domain includes both the mass losing star 
and the mass accreting star on the equatorial plane (Fig. \ref{fig_1}). 
The computational domain that contains the mass losing star has 
$20 \times 61$ grid points, 
while that contains the mass accreting star has $44 \times 61$ grid points. 
These two domains are connected by a narrow strip 
that has $3\times 19$ grid points. 

\subsection{NUMERICAL TREATMENT OF RADIATIVE TRANSFER}
It is assumed that energy discharge due to radiative cooling occurs 
in the vertical directions from the top and bottom surfaces of the disc. 
Because the present flowfield is confined 
in a two-dimensional equatorial plan, 
it is necessary to assume the distribution of physical quantities 
in the vertical direction in the disc for calculation of radiation. 
We assume a hydrostatic balance in the vertical direction 
to determine the thickness of the disc, 
and that the physical quantities 
in the vertical direction are constant with the local values 
in the equatorial plan. The thickness of the disc $H$ is given by
\begin{equation}
H=\frac{c_{s}}{\Omega_{\mbox{K}}}=
c_{s}\sqrt{\frac{r_{1}^{3}}{GM_{1}}}, 
\end{equation}
where $c_{s}$, $\Omega_{\mbox{K}}$, $r_{1}$ and $G$ are 
the sound speed of the gas on the disc, the Keplerian frequency, 
the radial distance from the mass accreting star and 
the gravitational constant, respectively. 

The coupling of the flowfield and the radiation field is occurred 
through the term of $\partial F_{z}/\partial z$ that appears 
in the energy equation. 
$\partial F_{z}/\partial z$ is obtained 
by solving the following equation for radiation energy 
\begin{equation}
\frac{\partial F_{z}}{\partial z}=4\pi \chi B-c \chi E, 
\label{eq-5}
\end{equation}
where $B$, $E$, $c$ and $\chi$ are the Planck function, 
the radiation energy density per unit volume, 
the speed of light and opacity, respectively. 
Because the speed of light is far larger than the sound velocity, 
and we only seek for a steady state solution, 
we ignore the time differential term that appears 
in the radiative transfer equation. 

In the present work we employ the Kramer's opacity 
which has no dependence on frequency. 
This Kramer's opacity is suitable 
for the calculation of pure hydrogen gas above $10^{4}[\mbox{K}]$, 
and is written in the following form \citep{CSW88,Stehle99}  
\begin{equation}
\chi=2.8 \times 10^{24}\rho^{2} T^{-3.5} [\mbox{cm}^{-1}],
\label{eq-6}
\end{equation}
where $T$ is the temperature of the gas. 

As mentioned earlier, we consider the radiative transfer only 
in the vertical direction of the disc, 
i.e. the $z$ direction perpendicular to the $x\mbox{-}y$ plane. 
The radiative transfer within the disc is ignored. 
Moreover, the radiative transfer is taken into account 
in the calculations only in the computational domain 
that contains the mass accreting star and hence the accretion disc. 

The radiative heat flux in the accretion disc is calculated 
by the FLD approximation \citep{AW74,LP81}. 
We assume the gas is in local thermodynamic equilibrium  
at a temperature that needs not correspond to 
that of the radiation field. 
This is because we solve the radiation energy equation (\ref{eq-5}) 
to determine the radiation energy density $E$. 
With the FLD approximation, 
the radiative flux can be written in the form of Fick's law 
of diffusion \citep{LP81} as
\begin{equation}
F_{z}=-D\frac{\partial E}{\partial z},
\label{eq-7}
\end{equation}
with a diffusion coefficient, $D$, given by
\begin{equation}
D=\frac{c \lambda}{\chi}.
\label{eq-8}
\end{equation}
The dimensionless function $\lambda=\lambda(E)$ is called 
as the flux limiter. We employ Minerbo's model \citep{Minerbo78} 
as a choice of $\lambda$, i.e. a constraint on the anisotropy 
of the radiation field. 
Minerbo assumed a piecewise linear variation of the specific intensity 
with angle, and found the functional form of the flux limiter as
\begin{equation}
\lambda(R)=\left\{
\begin{array}{ll}
2/(3+\sqrt{9+12R^{2}}) & \quad \mbox{if} \quad 0 \le R \le 3/2, \\
(1+R+\sqrt{1+2R})^{-1} & \quad \mbox{if} \quad 3/2 < R < \infty,  
\end{array}
\right.
\label{eq-9}
\end{equation}
where $R$ is a dimensionless quantity 
$R=\left| \partial E/\partial z \right|/(\chi E)$. 
In the optically thin limit $R \rightarrow \infty$, 
the flux limiter gives 
\begin{equation}
\lim_{R \to \infty} \lambda(R)=\frac{1}{R}, 
\label{eq-10}
\end{equation}
to the first order in $R^{-1}$. 
The magnitude of the flux therefore approaches 
$\left|F_{z}\right|= c\left| \partial E/\partial z\right|/(\chi R)=cE$, 
which obeys the causality constraint. 
In the optically thick or diffusion limit $R \rightarrow 0$, 
the flux limiter gives 
\begin{equation}
\lim_{R \to 0} \lambda(R)=\frac{1}{3}, 
\label{eq-11}
\end{equation}
to the first order in $R$, so that the flux takes 
the value given by equation (\ref{eq-7}).
 
The method of the solving the radiation energy equation (\ref{eq-5}) 
is described in Appendix \ref{appendixa}. 

\section{RESULTS}\label{section3}

\begin{figure*}
\vspace{15cm}
\caption{Density contours for each case. 
           (a), (b) and (c) show the density contours for the cases of 
           $\gamma=5/3$, $1.2$ and $1.01$ not accounting for 
           the radiative cooling, respectively. 
           (d) shows the density contours for the case of $\gamma=5/3$ 
           accounting for the radiative cooling. 
           The density range in (a), (b), (c) and (d) is 
           from $1.0 \times 10^{-13}$ to $1.0 \times 10^{-12}$, 
           from $1.7 \times 10^{-13}$ to $2.6 \times 10^{-12}$,  
           from $5.4 \times 10^{-13}$ to $1.1 \times 10^{-11}$ and 
           from $2.5 \times 10^{-13}$ to 
           $4.9 \times 10^{-12}[\mbox{g cm}^{-3}]$, respectively. 
           Each density range 
           is divided by $20$ lines with an equal increment. 
           The system rotates counterclockwise. 
           }
\includegraphics{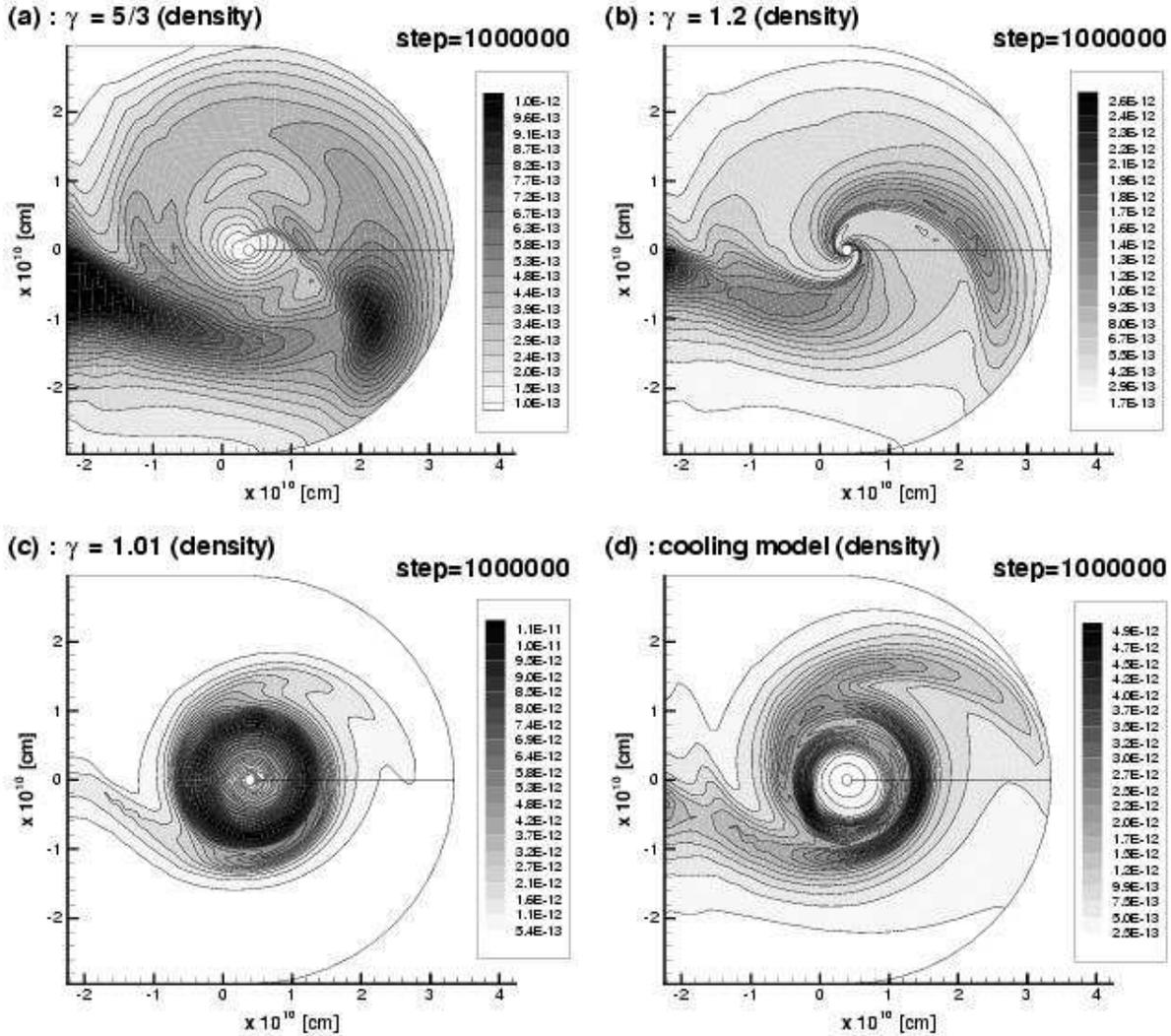}
\label{fig_2}
\end{figure*}

\begin{figure*}
\vspace{15cm}
  \caption{Temperature contours with logarithmic scale for each case. 
           (a), (b) and (c) show the temperature contours for the cases of 
           $\gamma=5/3$, $1.2$ and $1.01$ not accounting for 
           the radiative cooling, respectively. 
           (d) shows the temperature contours for the case of $\gamma=5/3$ 
           accounting for the radiative cooling. 
           The temperature range in (a), (b), (c) and (d) is 
           from $\log_{10} T[\mbox{K}]=5.9$ to $8.1$, 
           from $5.9$ to $7.5$, 
           from $4.6$ to $7.2$ and  
           from $4.6$ to $7.2$, respectively. 
           The temperature range is divided by $20$ 
           lines with an equal increment. 
           }
\includegraphics{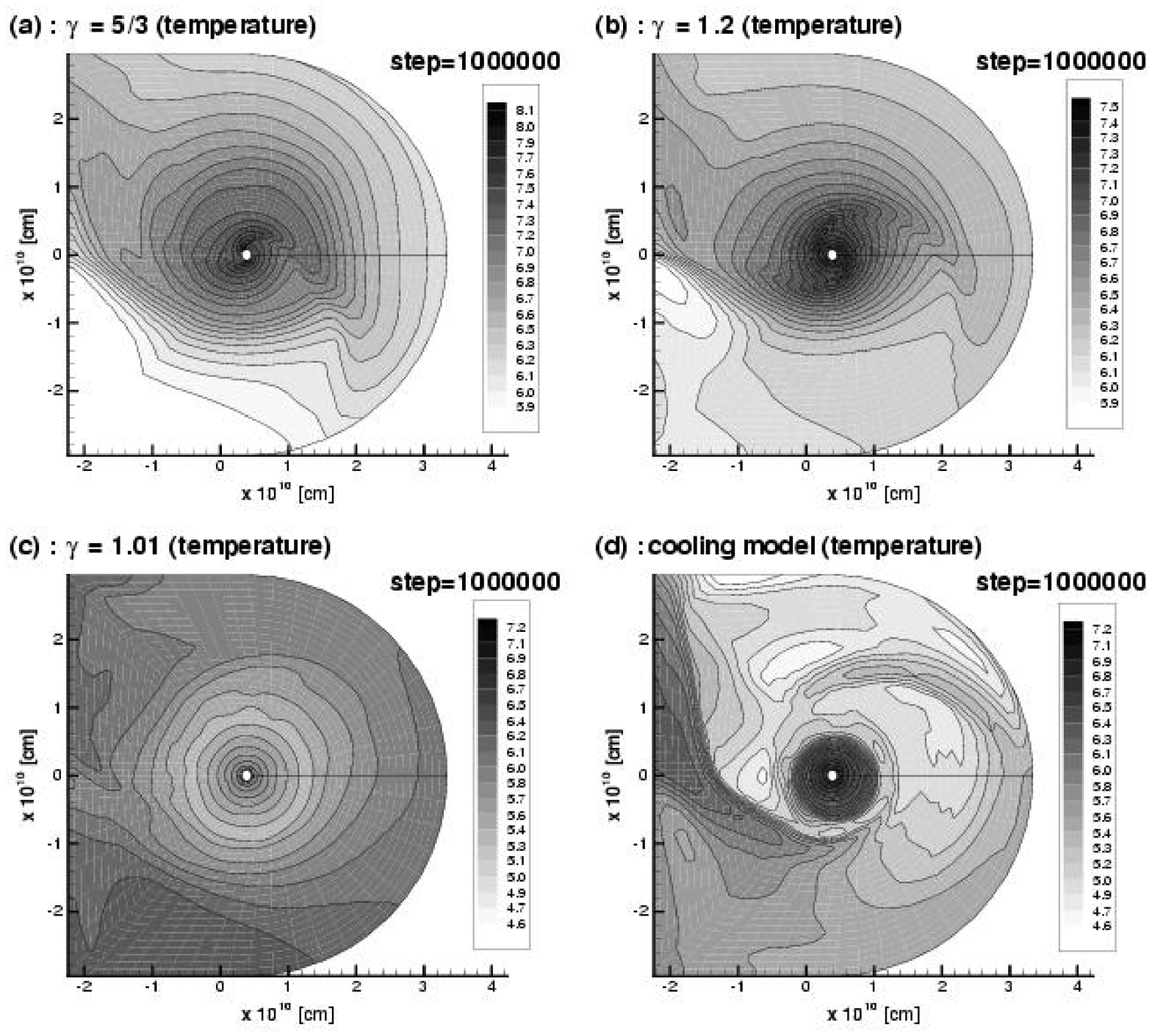}
\label{fig_3}
\end{figure*}

Figs. \ref{fig_2} (a), (b) and (c) show the density distribution 
in a (quasi) steady state (at $10^{6}$ step) 
for the cases of adiabatic gas 
with $\gamma=5/3$, $1.2$ and $1.01$, respectively. 
In these calculations radiative cooling effect is not included. 
It is shown that the disc structure is formed and 
the spiral shocks are appeared in each case. 
For the case of $\gamma=5/3$, 
the spiral arms are less clear as shown in Fig. \ref{fig_2} (a).
This is because the contour lines are plotted at equal interval 
while the density in the disc for this case is relatively lower 
than that in the L1 stream.

It should be noted that the steady state is not steady 
in a strict sense, but slightly oscillates. 
Especially, for the case of $\gamma=5/3$, 
the density pattern and the shock position periodically change with time, 
although the general pattern in the inner and outer region 
remains unchanged. 
The temporal fluctuation of the disc structure, however, is not 
significant and the density distribution shown in Fig. \ref{fig_2} (a) 
gives a typical one for the case of $\gamma=5/3$. 
This is suggested by the fact that the present density distribution 
virtually coincides with that of the ensemble average taken 
from $2.0 \times 10^{5}$ to $1.0 \times 10^{6}$ time steps.

The pitch angle of spiral arms has a clear correlation with $\gamma$. 
The lower $\gamma$ thus leads to a tightly wound spiral shocks. 
These results are consistent 
with the previous works \citep{SMH86a,MMM00}. 

The maximum Mach number in the disc is about $4$, $5$ and $23$ 
for the cases of $\gamma=5/3, 1.2$ and $1.01$, respectively. 

Fig. \ref{fig_2} (d) shows the density distribution 
for the case in which the specific heat ratio is fixed as $5/3$ 
and the effect of radiative cooing is accounted for. 
Obviously, one can see that the disc structure is formed 
and spiral shocks are appeared. 

Spiral shocks disappear near the mass accreting star 
and nearly axisymmetric disc structure is formed. 
Those flow patterns are quite similar to that obtained 
in the calculation of adiabatic case with $\gamma=1.01$, 
although the absolute value of density in the disc becomes smaller. 

The maximum Mach number in the disc is found to be $23$ for this case 
which is the same as that for the case of $\gamma=1.01$. 

Figs. \ref{fig_3} (a), (b) and (c) show the temperature distribution 
in a (quasi) steady state with logarithmic scale 
for the cases of adiabatic gas 
with $\gamma=5/3$, $1.2$ and $1.01$, respectively. 
As can be seen in Fig. \ref{fig_3} (a) and (b), 
the temperature of the gas in the disc is very high 
and becomes above $10^{7}[\mbox{K}]$ 
near the mass accreting star. 
The smaller $\gamma$ value for this case results 
in nearly isothermal distribution in the disc 
that gives substantially lower temperature 
near the mass accreting star. 
As shown in Fig. \ref{fig_3} (c), 
compared with the other adiabatic cases, 
the temperature distribution for the case of $\gamma=1.01$ 
is found to be close to flat.

Fig. \ref{fig_3} (d) shows the temperature distribution 
with logarithmic scale in a (quasi) steady state 
for the case of $\gamma=5/3$ with radiative cooling. 
Compared with the other adiabatic cases without radiative cooling, 
the temperature in the accretion disc is 
significantly lowered to less than $10^{5}[\mbox{K}]$, 
except for the small disc region near the mass accreting star 
where the temperature exceeds $10^{7}[\mbox{K}]$. 
In other words, the accretion disc exhibits a dual structure, 
i.e. the significantly cooled outer region and the inner core region 
where temperature is very high. 
The precise structure in the inner core region, 
however, is difficult to resolve with the present mesh system, 
because the pitch angle of the spiral shocks 
becomes smaller as approaches to the inner region. 

\begin{figure*}
\vspace{15cm}
\caption{The evolution of the total mass in the mass accreting region
           denoted by solar mass unit.
           The total mass is defined as a sum of all the masses 
           in the mass accreting region. 
           }
\includegraphics{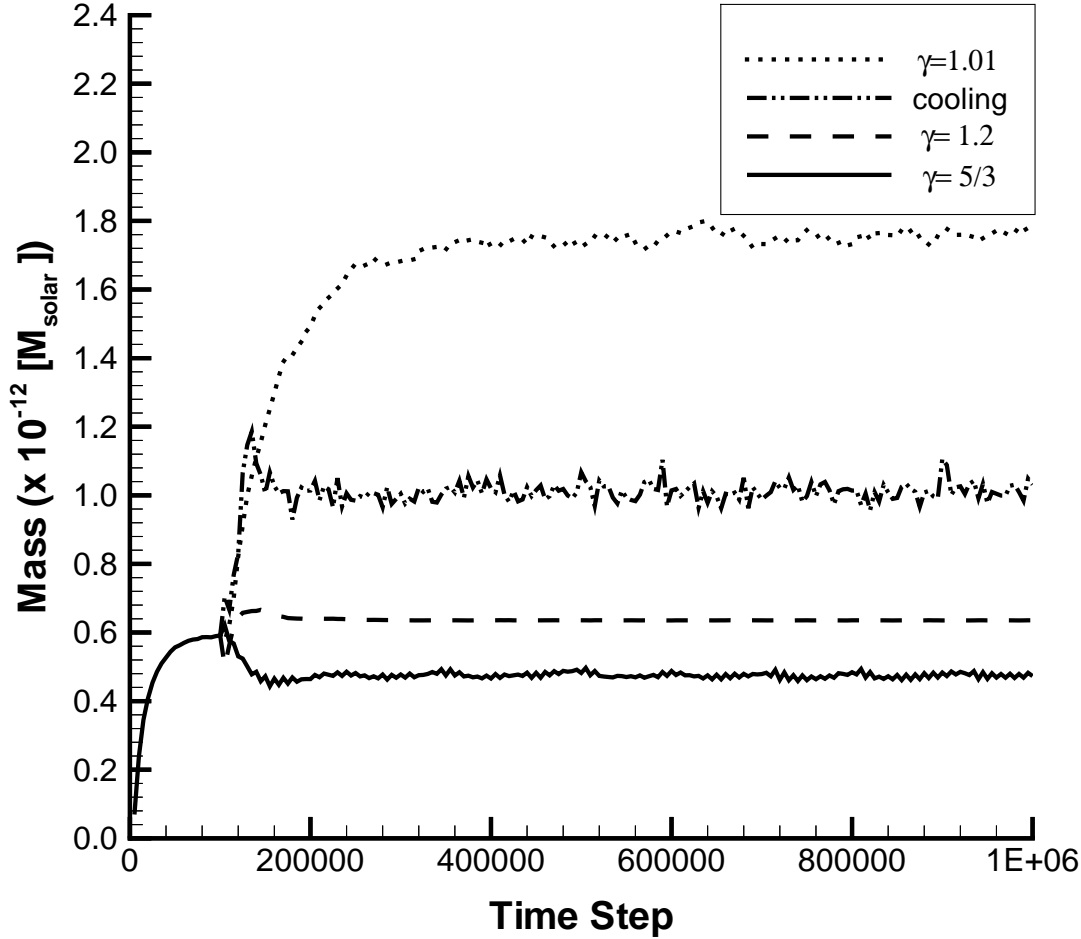}
\label{fig_4}
\end{figure*}

The evolution of the total mass in the mass accreting region is 
shown for each case in Fig. \ref{fig_4}. 
The total mass is defined as a sum of all the masses 
in the mass accreting region. 
The horizontal axis denotes the time step. 
$1.0 \times 10^{6}$ time steps correspond to more than 
$700$ orbital periods.  
In all the cases, the total mass in the mass accreting region becomes 
almost constant after $3.0 \times 10^{5}$ time steps 
but begin to oscillate slightly around each mean value. 
Therefore, we can say that each flowfield 
in the accretion disc reaches a (quasi) steady state. 
The total mass for the case accounting for radiative cooling is found 
to be larger than that for the case of $\gamma=1.2$, 
and smaller than that for the case of $\gamma=1.01$. 
From the results, we can say that the overall behavior of the flowfield 
as well as the evolution of the total mass of the adiabatic case 
with a smaller $\gamma$ value is quite similar to the one 
with radiative cooling shown in the present study. 
Although the precise $\gamma$ value can change for case-to-case, 
one can conclude that the use of a lower $\gamma$ value 
for simulating radiative cooling effect is almost a valid assumption. 

\begin{figure*}
\vspace{15cm}
\caption{Density contours for the case of $\gamma=5/3$ 
           accounting for the radiative cooling 
           for the different parameters of $\rho_{0}$ and $c_{0}$. 
           (a), (b), (c) and (d) show the density contour for the cases 
           of $(\rho_{0},c_{0})
           =(0.5 \times 10^{-9}, 48),
           (0.5 \times 10^{-8}, 73),
           (1.0 \times 10^{-7}, 24)$ 
           and 
           $(0.5 \times 10^{-9}[\mbox{g cm}^{-3}], 73[\mbox{km s}^{-1}])$, 
           respectively. 
           The figure of (d) is same as that of Fig. \ref{fig_2} (d), 
           which is the standard case. 
           The density range in (a), (b) and (c) is 
           from $2.5 \times 10^{-13}$ to 
           $4.9 \times 10^{-12}$, 
           from $2.4 \times 10^{-12}$ to 
           $4.4 \times 10^{-11}$ and 
           from $4.7 \times 10^{-12}$ to 
           $8.4 \times 10^{-11}[\mbox{g cm}^{-3}]$, respectively. 
           The density range in (d) is same as it of (b). 
           Each density range 
           is divided by $20$ lines with an equal increment. 
           }
\includegraphics{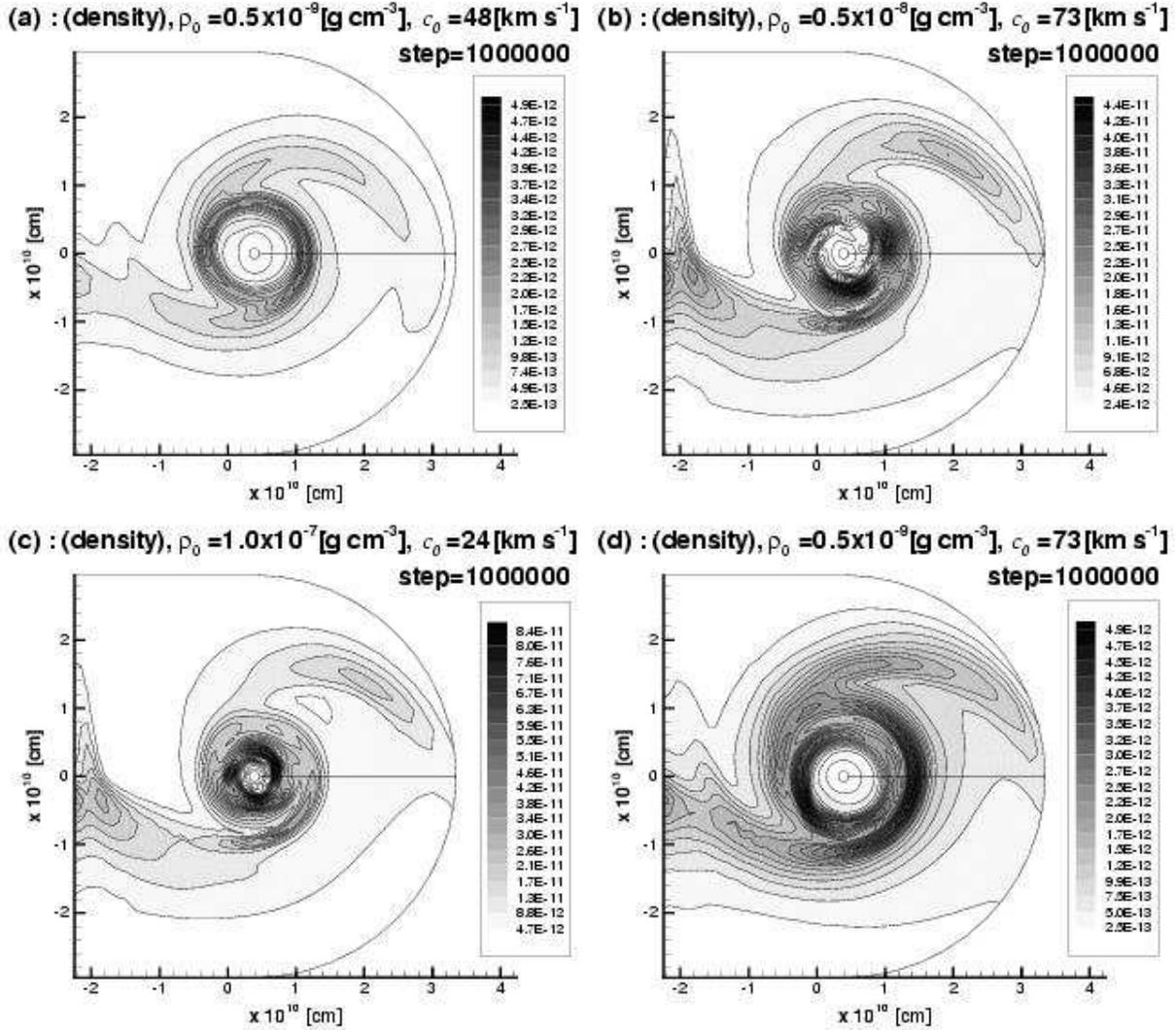}
\label{fig_5}
\end{figure*}

Finally, let us examine the influence of gas properties from 
the mass losing star. 
Because the process of discharge and absorption of radiation 
depends on density and temperature of the gas 
in the accretion disc that is influenced primarily 
by the gas from the mass losing star, 
we need to know what are the consequences of changing properties of the gas. 
Fig. \ref{fig_5} (a) shows the density contours in the accretion disc 
for the lower temperature case 
in which $\rho_{0}=0.5 \times 10^{-9}[\mbox{g cm}^{-3}]$ 
and $c_{0}=48[\mbox{km s}^{-1}]$. 
In the calculation, the radiative cooling effect is considered. 
In this case, the accretion disc slightly shrinks, 
but the overall feature is unchanged 
if compared with the standard case shown in Fig. \ref{fig_5} (d). 
On the other hand, if we increase the surface density 
of the mass losing star, some changes are seen in the results. 
Fig. \ref{fig_5} (b) shows the density contours for the case 
with $\rho_{0}=0.5 \times 10^{-8}[\mbox{g cm}^{-3}]$ 
and $c_{0}=73[\mbox{km s}^{-1}]$. 
Note that the density scale is changed to show the overall distribution. 
In this case, the inner disc region again appears 
where density becomes very low, but its radius becomes smaller. 
Surrounding this inner region, there appear several spiral arms 
in the intermediate region. 
This intermediate region has a ring like structure, 
which is connected to an outer spiral arm and also 
to the incoming flow from the L1point. 
In Fig. \ref{fig_5} (c), the density contours 
for the limiting case of $\rho_{0}=1.0 \times 10^{-7}[\mbox{g cm}^{-3}]$ 
and $c_{0}=24[\mbox{km s}^{-1}]$ are shown. 
One can see that the overall feature is quite similar to that shown 
in Fig. \ref{fig_5} (b), 
though the inner disc region further shrinks.

\section{CONCLUDING REMARKS}\label{section4}
Two-dimensional hydrodynamic calculations of an invisicid flowfield 
in a close binary system are carried out 
by solving the Euler equations with radiative transfer. 
In the present study, 
the specific heat ratio is assumed to be constant 
while radiative cooling effect is included as a non-adiabatic process. 
The cooling effect of the disc is considered by discharging energy 
in the vertical directions from the top and bottom surfaces of the disc. 
We use the flux-limited diffusion approximation 
to calculate the radiative heat flux values. 
The obtained results are summarized as follows:

\begin{description}
\item[(i)] Around the mass accreting star, 
there appears an accretion disc and spiral shocks are formed in the disc, 
even for the case with radiative cooling. 
\item[(ii)] The total mass of the disc reaches a (quasi) steady value. 
It becomes larger when radiative cooling effect is accounted for than 
that assuming $\gamma=5/3$, 
and is close to the value for the case of lower $\gamma$. 
\item[(iii)] The lower $\gamma$ value 
for an adiabatic gas which has been employed 
in the simulation of accretion discs in order to account 
for radiative cooling effect is found to be almost a valid assumption 
in terms of the overall flow features and also the amount of the total mass. 
\end{description}

What are shown in this work are the results 
of two-dimensional hydrodynamic calculation of inviscid flowfield accounting 
for radiative cooling effect. 
In order to obtain more rigorous results for the accreting flowfield, 
we need to explore three-dimensional flowfield that is coupled 
with radiative transfer. 
In such case, one needs to account for radiative transfer 
that is parallel to equatorial plane. 
This is because the accretion disc is not necessarily a flat thin disc. 
It is also needed to employ a more detailed radiation model 
that can consider spectral dependence of opacity. 
For such calculation, a flux-limited diffusion model is no more applicable 
and we need further 
to employ a detailed simulation method such as to use ray-tracing technique 
to solve radiative transfer equations. 
Our future studies will focus on these aspects.

\appendix

\section[]{Discretizing of Radiation Energy Equation}
\label{appendixa}
Let us consider the method of solving 
the radiation energy equation (\ref{eq-5}). 
Using the flux-limited diffusion approximation (\ref{eq-7}), 
equation (\ref{eq-5}) can be rewritten in the following form 
for radiation energy density $E$ as  
\begin{equation}
-\frac{\partial}{\partial z}\left(
D\frac{\partial E}{\partial z}\right)
=4 \pi \chi B - c \chi E. 
\label{eq-a1}
\end{equation}
Because the distribution of physical quantities 
in the vertical direction is assumed to be constant, 
the diffusion coefficient $D$ is independent of the height $z$. 
Therefore, we replace the partial differential in $z$ by multiplying $1/H$. 
As a result, equation (\ref{eq-a1}) is discretized as  
\begin{equation}
2 D^{n}\frac{E^{n+1}}{(H^{n})^{2}}=
4 \pi \chi^{n} B^{n} - c \chi^{n} E^{n+1}, 
\label{eq-a2}
\end{equation}
where the superscript $n$ denotes the time step. 
The factor $2$ in the left hand side of the equation (\ref{eq-a2}) 
appears because discharging energy by radiation occurs 
from the top and bottom surfaces of the disc. 
Except for the radiation energy density $E$, 
the dependent variables are evaluated at time step $n$. 
Equation (\ref{eq-a2}) is solved for $E^{n+1}$. 
This $E^{n+1}$ is substituted into the discrete form 
of $\partial F_{z}/\partial z$, i.e. 
$2 D^{n} E^{n+1}/(H^{n})^{2}$ which is further substituted 
into the energy equation of the Euler equations. 
This completes the coupling procedure of the flowfield with radiation.

\bsp

\label{lastpage}

\end{document}